\documentclass[showpacs,amsmath,amssymb,twocolumn]{revtex4}
\usepackage{graphicx}

\begin{document}

\title{Bistability, softening, and quenching of magnetic moments in
Ni-filled carbon nanotubes}

\author{Eduardo M. Diniz}
\email{diniz@fisica.ufmg.br}
\author{Ricardo W. Nunes }
\author{H\'elio Chacham}
\author{M\'ario S. C. Mazzoni}
\email{mazzoni@fisica.ufmg.br}

\affiliation{Departamento de F\'{\i}sica, ICEX, Universidade Federal
de Minas Gerais, CP 702, 30123-970, Belo Horizonte, MG, Brazil.}

\begin{abstract}
The authors apply first-principles calculations to investigate the
interplay between structural, electronic, and magnetic properties of
%
%
nanostructures composed of narrow nanotubes
filled with metallic nanowires. The focus is on the structural and
magnetic responses of Ni-filled nanotubes upon radial compression.
Interestingly, metastable flattened structures are identified, in
which radially deformed nanotubes are stabilized by the interactions
with the encapsulated wire. Moreover, our results
indicate a quenching of the magnetic moment of the wire upon
compression, as a result of the transfer of charge from the $s$ to the
$d$ orbitals of the atoms in the wire.
\end{abstract}


\maketitle

The possibility of metal encapsulation in carbon nanotubes has
attracted the attention of the physics community since the very
beginning of the research in this field. In the early years following
Iijima's paper on the synthesis of nanotubes \cite{iijima}, several
groups reported capillarity effects and filling of large diameter
multiwall nanotubes with a variety of elements
\cite{iijima2,tsang,ugarte}. However, the synthesis of sub-nanometer
structures of this type remained a challenge until recently, when a
protocol to fill very narrow nanotubes ($\sim$8.0\,\AA\ in diameter)
with metallic wires was reported \cite{terrones1,terrones2}.
Encapsulation protects the wire against oxidation, opening up the
possibility of the experimental confirmation of theoretical
predictions, such as the enhancement of magnetic moment of
transition-metal nanowires relative to bulk values
\cite{chang2,fu,seifert,jo}. In parallel to these reports, important
advances in manipulation processes have been achieved. For instance,
electric force microscopy measurements have been employed to probe
effects of deformation on the electronic and structural properties of
nanotubes, confirming theoretical predictions \cite{chacham,chang} of
a semiconductor-metal transition upon flattening of the nanotube cross
sections \cite{bernardo1,bernardo2}.

In the present work, we apply first-principles calculations to
investigate the behavior upon radial compression of narrow nanotubes
filled with metallic wires. The interplay between structural,
electronic, and magnetic properties is the main focus of our study.
We address the following questions: (i) How does the total energy
behave as the nanotube is flattened, when a metallic chain is
encapsulated within the nanotube, and how does this elastic response
compares with that of empty nanotubes~\cite{bernardo2}? (ii) If the
encapsulated metal is magnetic, how does the magnetic moment vary, if
it does at all, upon compression?
\begin{figure}[!ht]
\centering 
\includegraphics[width=7.0 cm]{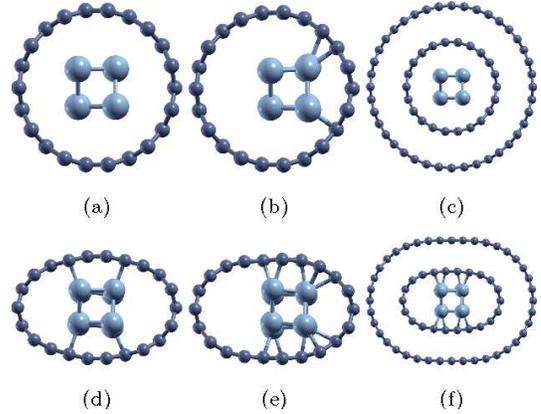}
\caption{(Color online) In (a), (b) and (c) are shown circular
cross-sections for on-center Ni$_8$@(11,0), off-center Ni$_8$@(11,0)
and on-center Ni$_8$@(11,0)@(20,0) configurations, respectively. In
(d), (e) and (f) we show unconstrained optimized flattened versions of
(a), (b) and (c), respectively.}
\label{geos}
\end{figure}

To address these and related questions we perform first-principles
electronic-structure calculations based on the Density Functional
Theory (DFT) \cite{dft} within the generalized gradient approximation
(GGA) \cite{gga,lda}. We employ the SIESTA implementation
\cite{siesta1,siesta2}, which makes use of norm-conserving
pseudopotentials \cite{pseudo1,pseudo2} and a basis set composed of
pseudo-atomic orbitals of finite range. The geometries in the
unconstrained calculations are relaxed until the total energy of the
unit cell changes by at most 0.03\,eV~\cite{tolerance}.

As a model for the metallic filling of the nanotube, we consider a Ni
nanowire (see Ref.~\cite{peng}) along the $z$ direction with four base
atoms at $(x,y,z)$ coordinates given by $(\pm b,0,0)$ and $(0,\pm
b,a/2)$, where $b = 1.52$\,\AA, and $a =2.36$\,\AA\ is the lattice
parameter along the $z$ direction. This choice is justified because,
in this geometry, the nanowire is nearly commensurate with a nanotube
with a diameter in the range of those observed in the above mentioned
experiments \cite{terrones1,terrones2}. From the initial geometry, the
whole nanowire-nanotube structure relaxes, and the nanowire undergoes
a small transformation (akin to a slight dimerization) into an
eight-atom unit that is commensurate with the nanotube period.

In Figs.~1(a) and 1(b), we show the cross sections of two metastable
energy-minima geometries (on-center and off-center, respectively) for
the encapsulation of the Ni wire in a (11,0) carbon nanotube. The
off-center configuration is found to be energetically favorable by
0.52\,eV/unit (i.e. 65\,meV/Ni-atom), when compared to the on-center
one. Such stability enhancement is followed by an increase of the
charge transfer from the wire to the nanotube (0.484\,$e$ in the
off-center as compared with 0.440\,$e$ for the on-center geometry),
indicating stronger interaction and hybridization. Concurrently, the
magnetic moments are 5.83\,$\mu_B$ and 4.28\,$\mu_B$ for the on-center
and off-center configurations, respectively. This suggests that there
is a reduction of $\mu$ caused by the enhanced C-Ni interaction, which
is in agreement with the findings of Jo et al.~\cite{jo}, who found
a decrease in the magnetic moments of encapsulated Ni wires when going
from the (7,7) nanotube to narrower (5,5) one. In the following, we
show that this is indeed the case, and ascribe this behavior to an
increase in the charge transfer from the $s$ to the $d$ orbitals of
the Ni atoms, upon confinement.

Taken together, the above results for the wire energetics and magnetic
moments suggest that radially compressed structures might be
relevant. To enforce radial compression, we use a methodology similar
to the one employed in the investigation of the elastic and electronic
properties of flattened carbon nanotubes \cite{chacham,bernardo1}:
deformed configurations are produced by building initial geometries
for the nanotubes with cross sections composed of semi-circles joined
together by straight lines. We define the compressive strain $\eta = 1
- d/D$, where $D$ is the diameter of the undeformed nanotube and $d$
is the distance between the straight lines, kept fixed during the
relaxation by constraining the positions of the atoms belonging to the
planar regions.

%
%
The results of our calculations for the radially compressed Ni-filled
nanotubes are displayed in Figs.~2(a) and (b), respectively for the
total energy ($\Delta E$) and magnetic moment ($\mu$) as a function of
$\eta$, for both the on-center (red squares) and off-center (black
circles) configurations. Fig.~2(a) shows metastable (local minima)
configurations around $\eta = 0.22$ and $\eta = 0.29$ for the
on-center structure, and around $\eta = 0.26$ for the off-center
case. At these minima, the magnetic moment is substantially reduced
both from its initial ($\eta = 0$) value and from the
free-standing-wire value of $\mu = 6.8\,\mu_B$. For the on-center
configuration, in which the initial interaction among C and Ni atoms
is weak, the magnetic moment steadily decreases (apart from small
fluctuations) with the deformation. In the off-center case, the larger
interaction substantially reduces the magnetic moment at the very
beginning of the process, and a further decrease is observed for
larger values of $\eta$.
\begin{figure}[!ht]
\centering
\includegraphics[width=7.0 cm]{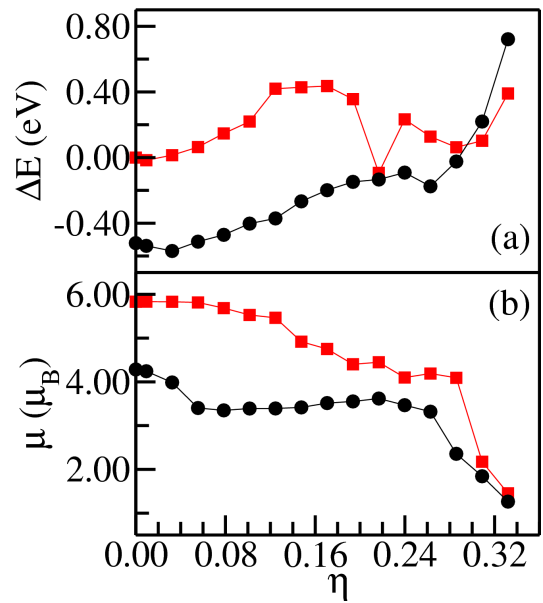}\label{energy-spin}
\caption{(Color online) (a) Energy and (b) total magnetic moment for
an 8-atom Ni wire inside a (11,0) carbon nanotube as a function of flattening
($\eta$ is defined in the text). Red squares and black circles
correspond to the on-center and off-center configurations,
respectively.}
\end{figure}

The minima for $\eta > 0$ of Fig.~2(a)
were obtained under the geometrical constraints
of the radial compression. We removed such constraints, and obtained
the unconstrained metastable energy minima shown in Figs.~1(d) and
(e). Interestingly, the self-flattened structure of Fig.~1(e) is the
second most stable in Fig.~1, only 0.35\,eV (0.04\,eV/Ni-atom) above that
of Fig.~1(b), which is the most stable one. This raises the
interesting possibility of achieving metastable flattened structures
of Ni-filled nanotubes through compression by the tip of an atomic
force microscope, for instance.
Further, the geometry shown in Fig.~1(e) presents an average C-Ni bond length
of 2.09\,\AA, which is in the range of typical C-Ni distances 
%
%
that have been reported in previous
calculations for
Ni interacting with carbon structures (2.07\,\AA\ to 2.20\,\AA)\cite{jo}.

\begin{figure}[!ht]
\centering
\includegraphics[width=7.0 cm]{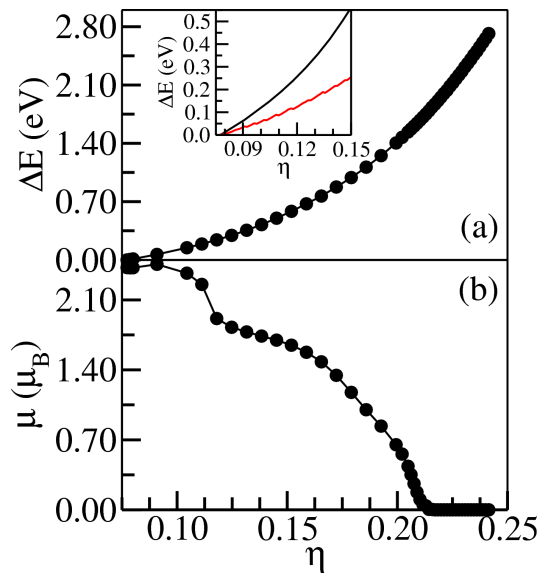}\label{zig9}
\caption{(Color online) (a) Energy and (b) total magnetic moment for
an 8-atom Ni wire inside a (9,0) carbon nanotube as a function of flattening
($\eta$ is defined in the text). In (a), the inset shows a comparison
between filled- (black line) and empty-nanotube (red line) responses to flattening}
\end{figure}

Previous modelling of the experimental results \cite{terrones2}
suggest that, for the diameter range that includes the (n,0) nanotubes
with $9 < n < 14$, a metallic wire diameter such as the one employed
in our models is consistent with the observed data \cite{terrones2}.
A particular situation occurs for the (9,0) structure, since the model
wire perfectly fits into the nanotube, making covalent bonds with the
walls even without deformation. In this case, the energy is a
monotonically increasing function of the flattening distance, and the
filled nanotube is harder than the empty one, as represented in the
plots of Fig. 3(a).  Interestingly, even in this case, the magnetic
moment is initially non-zero (2.40\,$\mu_B$), and decreases as a
function of the flattening parameter, reaching a null value for $\eta
= 0.21$ ($d = 5.79$\,\AA), as shown in Fig. 3(b). Notwithstanding the lower
formation energy per Ni atom for the filling of this nanotube
[-0.44~eV/atom versus -0.24~eV/atom for the (11,0) nanotube], the
kinematics of the process should be more favorable for the larger ones
($9 < n < 14$), since the Ni atoms tend to block the entrance of the
narrowest (9,0) nanotube, 
%
%
hindering the uniform filling of the tube.

Since the experimental results \cite{terrones1,terrones2} concern
double wall nanotubes, we have also considered the encapsulation of a
Ni wire inside a (11,0)@(20,0) structure. Fig.~1(f) shows the relaxed
geometry (without constraints) for an outer-tube flattening distance
of 13.1\,\AA, which resulted in a distance of 6.2\,\AA\ between
the inner-tube flat walls. We notice that practically all charge
transfers take place among the inner C atoms and the Ni atoms, the
magnetic moment decreases to 4.04~$\mu_B$, and the nanotube (both walls) mantains the
flat cross section, upon removal of the constraint.


In order to investigate whether the above results are more generally
valid, and thus not restricted to the above geometries, we study a
single Ni atomic chain encapsulated by the narrower (8,0) nanotube
(diameter of 6.43\,\AA). Again, we consider on-center and off-center
configurations. In this case, as shown in Fig.~4(a), we find two
off-center minima, which we call OFF1 and OFF2, corresponding to
off-center displacements between 0.50 and 0.87\,\AA\ in the OFF1, and
1.48 and 1.74\,\AA\ in the OFF2, depending on the degree of
flattening. The results presented in Fig.~4(b) indicate a sharp
decrease in the magnetic moment in these three metastable
configurations. Note that in the OFF2 case, the wire is close enough
to the nanotube wall for the magnetic moment to be totally suppressed
(or nearly so) by encapsulation, even without flattening. For the
on-center and OFF1 configurations, the magnetic moment is strongly
suppressed after a critical degree of flattening is reached.
\begin{figure}[!ht]
\centering
\includegraphics[width=7.0 cm]{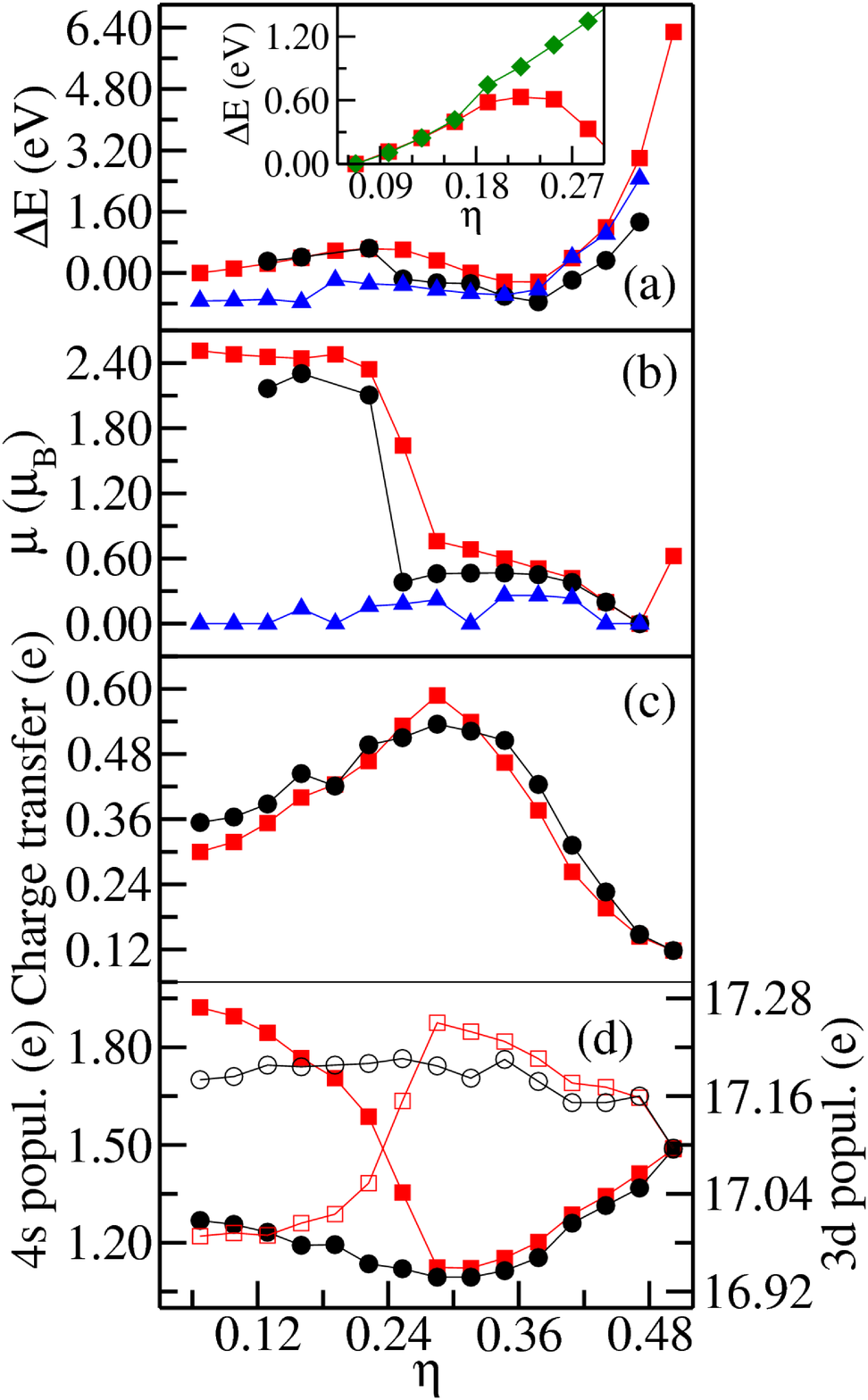}
\caption{Energy, magnetic moment, charge transfer and electronic
population for a linear Ni wire inside a carbon nanotube as a function
of flattening. Red squares, black circles, and blue triangles
correspond to the on-center, and the OFF1 and OFF2 off-center
configurations, respectively. In (a), the inset shows a comparison
between the empty- and filled-nanotube responses to flattening. In (d),
the filled symbols represent the $s-$population, while open symbols
represent the $d-$population.}\label{mono}
\end{figure}

The inset in Fig.~4(a) shows a comparison between the elastic
responses of empty and metal-filled nanotubes. We observe that for
$\eta > 0.16$, the metal-filled tube is softened by interaction with
the encapsulated wire, 
%
%
in contrast with the case of the (9,0) tube
filled with a thicker wire. 
The effect is even more dramatic for the OFF2 case, for which the
energy changes very mildly with flattening for $\eta\le 0.38$. As
mentioned before, the interpretation of these results may be pursued
by looking at the variations in charge transfer between the wire and
the tube, and also the intra-atomic charge transfer from $s$ to $d$
orbitals of the wire atoms. Figs.~4(c) and (d) show both charge
transfer mechanisms for the on-center and lowest-energy off-center
configurations as a function of $\eta$. It is clear that both are
maximum at the $\eta$ value where the strong quenching of magnetic
moment kicks in. Also, upon increasing the confinement, Ni
$s-$electrons are transferred to $d-$orbitals, quenching the magnetic
moment, as can be seen in Fig.~4(d). Basically, the same reasoning
applies to the wider wire-nanotube system discussed above.

In summary, we predict that metastable flattened configurations of
filled nanotubes may be achieved through radial deformation. Also,
the deformation induces a decrease in the magnetic moment, which may
cause a sharp transition to a non-magnetic state for sufficiently
narrow wires.

\begin{acknowledgments}
We acknowledge support from the Brazilian agencies CNPq, FAPEMIG, and
the projects Rede de Pesquisa em Nanotubos de Carbono, INCT de
Nanomateriais de Carbono, and Instituto do Mil\^enio em
Nanotecnologia-MCT.
\end{acknowledgments}


\begin{thebibliography}{100}
\bibitem{lda} The local density approximation (LDA) was tested as
well, and the results remained essentially unaltered.
\end{thebibliography}
\end{document}